\documentclass[notitlepage,aps,pra,twocolumn,groupaddress,superscriptaddress,10pt]{revtex4-1}

\usepackage{stmaryrd}
\usepackage{amssymb,amsmath,amsthm,amsfonts,amsbsy}
\usepackage{bm,bibunits,color,chngcntr,epsfig,epstopdf,graphicx,dsfont}
\usepackage{hyperref,lipsum,,makecell,mathrsfs,rotating}
\usepackage[english]{babel}
\usepackage[normalem]{ulem}
\usepackage[absolute]{textpos}

\newcommand{\be}{\begin{equation}}
\newcommand{\ee}{\end{equation}}
\newcommand{\bea}{\begin{eqnarray}}
\newcommand{\eea}{\end{eqnarray}}
\newcommand{\ket}{\rangle}
\newcommand{\bra}{\langle}

\newcommand{\I}{\mathds{1}}

\def\C#1{\mathcal #1}

\definecolor{gray}{gray}{0.9}

\usepackage{setspace}

\begin{document}
\newtheorem{theorem}{Theorem}
\newtheorem{prop}[theorem]{Proposition}
\newtheorem{corollary}[theorem]{Corollary}
\newtheorem{open problem}[theorem]{Open Problem}
\newtheorem{conjecture}[theorem]{Conjecture}
\newtheorem{definition}{Definition}
\newtheorem{remark}{Remark}
\newtheorem{example}{Example}
\newtheorem{task}{Task}

\title{Distributed quantum computing with black-box subroutines}

\author{Xiang Xu}
\affiliation{Institute of Theoretical Physics, Chinese Academy of Sciences, Beijing 100190, China \\
School of Physical Sciences, University of Chinese Academy of Sciences, Beijing 100049, China}
\author{Yuan-Dong Liu}
\affiliation{Institute of Theoretical Physics, Chinese Academy of Sciences, Beijing 100190, China \\
School of Physical Sciences, University of Chinese Academy of Sciences, Beijing 100049, China}
\author{Sha~Shi}
\affiliation{School of Telecommunication Engineering, Xidian University, Xi’an, Shann Xi 710071, China \\
Guangzhou Institute of Technology, Xidian University, Guangzhou 510555, China \\ 
Hangzhou Institute of Technology, Xidian University, Hangzhou, Zhejiang 311231, China
}
\author{Yun-Jiang~Wang}
\affiliation{School of Telecommunication Engineering, Xidian University, Xi’an, Shann Xi 710071, China \\
Guangzhou Institute of Technology, Xidian University, Guangzhou 510555, China \\ 
Hangzhou Institute of Technology, Xidian University, Hangzhou, Zhejiang 311231, China
}
\author{Dong-Sheng Wang}
\email{wds@itp.ac.cn}
\affiliation{Institute of Theoretical Physics, Chinese Academy of Sciences, Beijing 100190, China \\
School of Physical Sciences, University of Chinese Academy of Sciences, Beijing 100049, China}


\date{\today}
\begin{abstract}
In this work, we propose a general protocol for distributed quantum computing 
that accommodates arbitrary unknown subroutines.
It can be applied to scale up quantum computing through multi-chip interconnection, 
as well as to tasks such as estimating unknown parameters or processes for circuit depth reduction 
and constructing secure quantum cryptographic protocols.
Our protocol builds upon a few techniques we develop, 
such as the oblivious quantum teleportation and control, 
which can circumvent quantum no-go theorems on the manipulation of unknown objects.
Furthermore, 
we demonstrate that this protocol can be physically implemented 
using currently available quantum computing platforms.
These results suggest that our framework could provide a foundation 
for developing more advanced quantum algorithms and protocols in the future.
\end{abstract}

\maketitle

\begin{spacing}{1.2}

\section{Introduction}\label{sec:intro}






Since the initial establishment of universal quantum computing criteria more than twenty years ago, the field has witnessed remarkable advancements in experimental quantum computing platforms~\cite{LJL+10}. 
Currently, substantial challenges still exist for improving the quality of logical qubits,
scaling up multi-qubit quantum chips, 
and developing more superior quantum algorithms and protocols. 

The primary universal quantum computing model is the circuit model,
in which a quantum computing task often involves the preparation of initial states, 
the execution of a sequence of quantum gates, and the final measurements for readout~\cite{NC00}. 
Depending on how these steps are realized,
more diverse models and protocols have been developed, 
expanding the application scenario of quantum computing~\cite{W24rev}. 
To name a few, there are 
distributed quantum computing which can combine the computing power 
of a few smaller quantum chips or servers~\cite{CAF24,BCD25}, 
blind quantum computing which can realize a user's private task 
by a remote quantum server~\cite{BFK09},
and quantum von Neumann architecture which has a higher-level of integration 
of modular units~\cite{NC97,W20_choi,YRC20,W24_qvn}. 
These advanced architectural paradigms provide both 
the theoretical foundation and practical motivation for our current investigation.

The focus of our study is distributed quantum computing, 
which is believed to be a vital path to scale up quantum chips~\cite{CAF24,BCD25}.
There are inspiring progresses in recent 
years~\cite{ACR21,QLX24,TLW23,MGD23,Hwang_2025,PS23,UPR23,LMH23,JYL25,AYH25,GPA24,AFM24,LZFD24}, 
including distributed quantum algorithms~\cite{ACR21,QLX24,TLW23,MGD23,Hwang_2025},
techniques to break circuits into smaller blocks such as the circuit knitting~\cite{PS23,UPR23,LMH23}, 
and also experimental efforts to develop the interface among 
different types of qubits~\cite{JYL25,AYH25}.
However, to the best of our knowledge,
there are two important aspects which have not been well studied.
First, compared with the usual circuit model describing a single chip,
a massive amount of communication is required,
and generic strategies to reduce the communication cost are less explored.
Second, the remote control or execution of unknown subroutines 
is necessary and sometimes unavoidable, 
and this has not been studied for distributed quantum computing.

The paradigm of computation with ignorance is not only widespread but also fundamentally important. 
For instance, in estimation and learning tasks~\cite{Hay17}, 
an unknown parameter or process exists \emph{a priori},
and a computing scheme needs to be able to manipulate it. 
For the design of computing architecture and programming~\cite{HH13},
programs shall be stored and loaded on-demand as subroutines,
and there is no need to decode a program in order to load it.
In cryptography and secure multi-party computation~\cite{KL20}, 
tasks often require protecting sensitive information from dishonest participants.

In literature, there is another term that describes such a feature of ignorance,
which is ``oblivious.''
Still, there are different usages of this term and here we classify 
two types: the weak and strong notions.
For the weak notion, 
a quantum operation or protocol is considered oblivious when its execution does
not depend on specific knowledge of its operand. 
Examples include 
the oblivious amplitude amplification algorithm~\cite{BCC+14}
and an approach of oblivious logical wire for measurement-based quantum computing~\cite{SWP+17}.
The strong notion would tie obliviousness to security as in  
the seminal protocol of oblivious transfer~\cite{KL20}, 
where a server (Bob) transmits two bit strings to a user (Alice), 
who can retrieve only one of them while Bob remains unaware of Alice’s choice.
In this work, we employ the weak notion of obliviousness,
but will also show a potential connection with the strong one.


It is not easy to achieve obliviousness for quantum computing,
even the weak type.
A few no-go theorems~\cite{WZ82,Die82,May97,LC97,NC97,AFC14,OGH16,TMV18} exist for various tasks,
but fortunately, there are strategies to circumvent them.
Using quantum meanings for oblivious transfer is not secure~\cite{May97,LC97}, 
while under reasonable assumptions, such as the noisy storage model~\cite{WST08},
it can be information-theoretically secure.  
To use stored quantum program states, 
it was shown that~\cite{NC97} the program state has to be classical,
otherwise, an unknown program cannot be downloaded, 
analogous to the no-cloning theorem~\cite{WZ82,Die82}. 
Also different from classical data, 
unknown quantum states or gates cannot be added together~\cite{AFC14,OGH16,TMV18},
due to the unphysical meaning of global phase of a state or gate. 
Here we propose the oblivious quantum teleportation
and control schemes that improve previous bypass strategies~\cite{W22_qvn}.

In this work, we develop a universal distributed quantum computing protocol
with black-box subroutines based on a few primary oblivious quantum schemes. 
Our protocol has a few features. 
First and foremost, it is oblivious hence it allows application settings with unknown information.
For instance, it can be used to construct secure protocols.
Second, it allows a high level of integration and quantum programming,
which in general employs quantum operations to manipulate quantum operations,
using the theoretical framework of quantum superchannel~\cite{CDP08a}.
Third, the remote execution of black-box subroutines also reduce the communication cost
as there is no need to upload or download the information encoded in the subroutines.
Forth, it offers a flexible way to change the circuit depth 
by using a space-time tradeoff to realize teleportation.
Therefore, we believe our protocol can be implemented 
on current quantum chips 
and also used for developing more advanced quantum algorithms.

In Section~\ref{sec:ope} we start from a primary task for 
the storage of quantum program and its oblivious execution scheme.
In Section~\ref{sec:oqt} we present oblivious quantum teleportation,
and in Section~\ref{sec:oqc} we present oblivious quantum control scheme.
In Section~\ref{sec:oqa} we study a few more oblivious quantum algorithms, 
which serve as foundational building blocks for advanced quantum algorithms.
In Section~\ref{sec:phys} we study physical realization in the current leading platforms. 
In Section~\ref{sec:dqc} we present our protocol for distributed quantum computing
and compare with a few related protocols.
We conclude by a discussion of the implication of obliviousness to hardware architecture, 
software development, and application protocols. 


\section{Oblivious program execution}
\label{sec:ope}

In this section, we start from a primary oblivious quantum operation,
which is to obliviously load a quantum program.
This is based on Refs.~\cite{W20_choi,W22_qvn,LWLW23},
and in this work we refer to this task as oblivious program execution (OPE).

Quantum processes are described as 
completely positive, trace-preserving (CPTP) maps~\cite{Kra83} of the form
\be \C E(\rho)= \sum_i K_i \rho K_i^\dagger, \ee
also known as quantum channels, 
for $K_i$ known as Kraus operators
and input states $\forall \rho \in \C D(\C H)$ for a system $\C H$.
This includes unitary evolution and measurements as special cases,
which are the two essential parts for constructing quantum algorithms.
The Stinespring dilation theorem shows that 
a quantum channel can be realized by a unitary circuit $U$ with 
the final partial trace over an ancilla, and 
\be K_i= \bra i|U|0\ket \ee 
for $\{|i\ket\}$ denoting an orthonormal basis of the ancilla. 
Therefore, we can focus on the unitary case for simplicity.

In the usual setting, a quantum computing process or algorithm 
includes the preparation, evolution, and measurement steps,
and also possibly transmission of quantum states.
The main part of a quantum algorithm, or called a program,
is a unitary circuit.
Besides processing, the storage of programs is also essential.
Currently, there are mainly three types. 
A circuit $U$ can be stored as a classical description $[U]$,
which could be the description of the gate sequence contained in $U$.
It can also be stored as a hardware, denoted as $H(U)$,
which particularly suits the photonic platform. 
This form is not scalable, however,
as will be studied in details in Sec.~\ref{sec:phys}.

\begin{figure}
    \centering
    \includegraphics[width=0.3\linewidth]{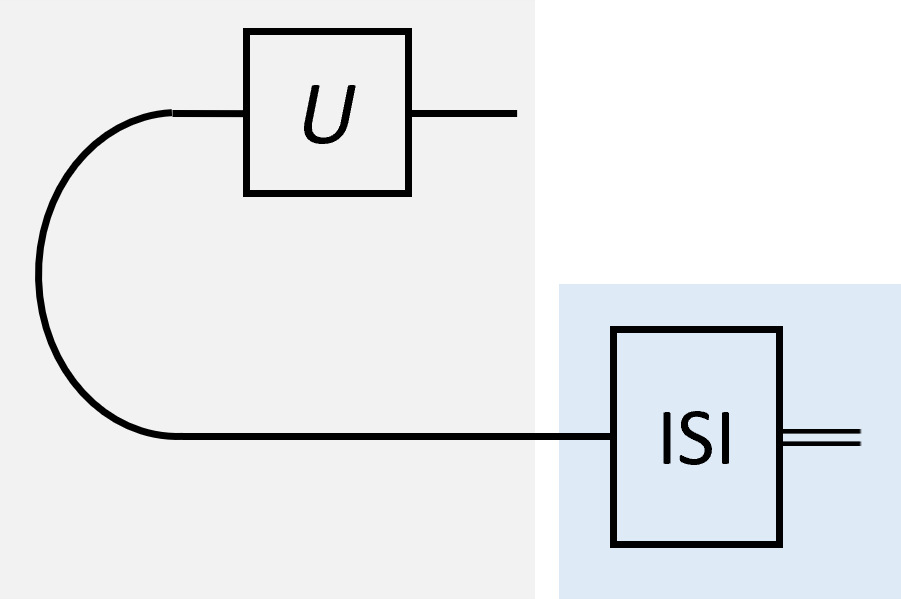}
    \caption{The Choi program state and the initial-state injection (ISI) scheme. 
    The curve represents a Bell state, i.e., ebit.}
    \label{fig:isi}
\end{figure}

Given $[U]$, the circuit $U$ can be executed, obviously. 
This is the opposite of being oblivious.
The third type is the Choi state 
\be |U\ket=(U\otimes \I)|\omega\ket, \ee 
according to the channel-state duality~\cite{Cho75,Jam72} 
for the Bell state or `ebit' as
$|\omega\ket:= \sum_i |ii\ket /\sqrt{d}$, $d=\text{dim}(\C H)$.
The form $|U\ket$ allows OPE via a measurement-based scheme.
Namely, it supports the write of initial state and 
the readout of observable. 
This also extends to the mixed-state case, shown in Appendix~\ref{sec:superc}.
For simplicity, denote the initial state as $|0\ket$,
and the goal is to obtain the probabilities 
\be p_a=|\bra a|U|0\ket|^2 \ee
for $\{|a\ket \bra a|\}$ 
as the projective measurement for readout. 
Here, the channel-state duality is involved 
to inject the initial state $|0\ket$
by a binary measurement $\{P_0, P_{\bar{0}}\}$ 
for $P_{\bar{0}}=\I - P_0$, and $P_0=|0\ket\bra 0|$.
Both outcomes work: 
for the case of $P_0$, $p_a$ is obtained;
for the case of $P_{\bar{0}}$, $p_a'=1-p_a$ is obtained.
This binary projective measurement is referred to as the 
initial-state injection (ISI) scheme~\cite{W20_choi}, see Fig.~\ref{fig:isi}.
Despite the simplicity of it,
the ISI scheme also includes an indirect binary Bell measurement as a special case, 
which plays the central role for the construction 
of oblivious quantum teleportation in Sec.~\ref{sec:oqt}.
The physical realization of ISI will be discussed in Sec.~\ref{sec:phys}.

We recall that the Choi-state program was motivated by the no-programming theorem~\cite{NC97},
according to which, 
for any possible form of quantum program state $|p_U\ket$, the gate $U$ cannot be executed as 
\be G |\psi\ket |p_U\ket = U|\psi\ket |p_U'\ket, \label{eq:cpu}\ee
for $G$ independent of $|\psi\ket$ and $|p_U\ket$.
The reason is that $\bra p_V| p_U \ket =0,$ $\forall U\neq V$,
i.e., the states $|p_U\ket$ are all orthogonal hence classical (e.g., $[U]$).
The encoding $|p_U\ket$ of $U$ acts like an encryption, 
and the downloading of $U$ is physically equivalent to 
a decryption and cloning of it~\cite{YRC20,W22_qvn}. 
We bypass this no-go result by requiring a weaker condition that 
only observation on $U|\psi\ket$ is needed, 
instead of the whole state. 
After all, it is the measurement outcomes that are needed. 
This does not sacrifice universality, 
since it can be equivalently formalized for gate simulation, 
state generation, or observable measurement~\cite{W15_QS}.




The state $|U^t\ket$ can also be obtained by exchanging the two ports of $|U\ket$
due to the vectorization property 
\be (A\otimes \I) |\omega\ket = (\I\otimes A^t) |\omega\ket \ee
for any operator $A\in \C B(\C H)$.
Given an unknown $U$, 
it is not easy to obtain $U^\dagger$, however~\cite{YSM23,MZC25}.
The simplest scheme is to prepare both $|U\ket$ and $|U^\dagger\ket$ on the first hand. 
Beyond that, we can employ the rebit approach~\cite{Aha03} to convert $U$ into an orthogonal operator
\be Q=\begin{pmatrix}
    U_1 & -U_2 \\ U_2 & U_1
\end{pmatrix}, \ee 
and convert input state $|\psi\ket$ into $|\Phi\ket=|R\ket |0\ket + |I\ket |1\ket$
for the real-imaginary part separation $U=U_1+iU_2$ and $|\psi\ket=|R\ket + i |I\ket$.
For any observable $A=\sum_a A_a|a\ket \bra a|$, the measurement probability 
on the final state is
\bea
p_a&=&\text{tr}\left((|a\ket \bra a| \otimes \I )Q|\Phi\ket \bra \Phi | Q^{\dagger}\right) \\
&=& \text{tr}\left((|a\ket \bra a|) U|\psi\ket \bra \psi | U^{\dagger}\right),
\eea 
that is, the statistics is the same as the original one. 
This will also be used to construct the quantum control unit in Sec.~\ref{sec:oqc}.

\begin{figure}
    \centering
    \includegraphics[width=0.7\linewidth]{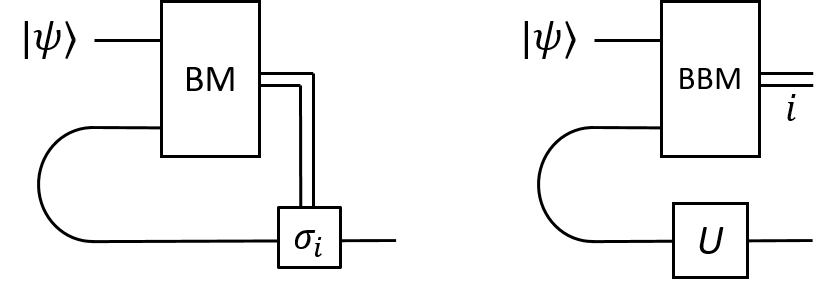}
    \caption{The standard quantum teleportation by Bell measurement (BM) (left)
    and the oblivious quantum teleportation by the indirect binary Bell measurement (BBM) (right).}
    \label{fig:tele}
\end{figure}

\section{Oblivious quantum teleportation}
\label{sec:oqt}

In this section, we introduce the oblivious quantum teleportation (OQT),
which can achieve the oblivious composition of programs $U_n \cdots U_2 U_1$.
This is an alternative of the previous universal quantum teleportation~\cite{W20_choi}, 
which is not oblivious,
and can also be viewed as a special ISI scheme 
but plays a distinct role. 

In the standard quantum teleportation (QT)~\cite{BBC+93},
an unknown state $|\psi\ket$ of a system S is teleported to another system B
by the Bell measurement $M_\text{AS}(i)$
according to
\be |\psi\ket_\text{B}= \sigma_{i,\text{B}}M_\text{AS}(i)
|\omega\ket_\text{AB}|\psi\ket_\text{S},  \ee 
with Pauli byproducts $\sigma_{i,\text{B}}$ being corrected
and consuming an ebit $|\omega\ket_\text{AB}$ as resource.
See Fig.~\ref{fig:tele}(left).
In a certain sense, it is oblivious since the teleportation circuit 
(i.e. Bell measurement) is independent of the input state $|\psi\ket$.

For quantum gate teleportation~\cite{GC99}, however, 
the scheme depends on the teleported gate $U$, i.e., 
it is not oblivious.
The original approach is to classify gates as the Clifford hierarchy 
\be \C C_n=\{U: UPU^\dagger \in \C C_{n-1} \} \ee 
for all $P$ in the Pauli group $\C C_1$,
and use lower-level gates to realize higher-level gates,
while the Pauli byproduct depends on the teleported gate.
For instance, in the setting of stabilizer codes
the so-called $T$ gate can be realized by gate teleportation~\cite{BK05}.
Recently, by using the large symmetry in teleportation
a universal gate teleportation scheme is found~\cite{W20_choi},
for which the byproduct depends on the teleported gate,
and the byproduct are not Pauli operators anymore.
In the light of the no-programming theorem~\cite{NC97},
gate teleportation is possible since 
the teleportation depends on an input program $U$.

We can achieve oblivious gate teleportation by a slight 
modification of Bell measurement.
The idea is to construct a special type of ISI scheme 
with the desired injected state as $|\omega\ket$.
Namely, by grouping the Pauli byproduct into two sets:
the trivial one $\sigma_0=\I$ and the nontrivial set $\{\sigma_i\}$ (for $i\neq 0$),
and use a Toffoli-type gate to extract the parity information 
of being trivial or nontrivial to a qubit ancilla,
the projective measurement on which with
\be P_0=|\omega\ket\bra\omega|,\; P_1=\I-|\omega\ket\bra\omega| \ee 
will realize either an identity or a channel $\C P$ formed by equal-weight $\sigma_i$.
There is no need to correct the Pauli byproduct.
See Fig.~\ref{fig:tele}(right).
The channel $\C P$ relates to the completely depolarizing channel $\Delta$~\cite{NC00}
by deleting its identity operator.
Then it is clear to see the oblivious teleportation of $U$ onto input $|\psi\ket$
realizes two outcomes 
\be P_0: U|\psi\ket; \; P_1: \frac{1}{d^2-1}\left(d\I - U\psi U^\dagger\right), \ee 
for $\psi:=|\psi\ket\bra\psi|$.
This also easily extends to the mixed state case.
The offset due to $\I$ for the outcome $P_1$ can be easily dealt with
for observable measurement.

More importantly, it extends to a sequence of teleportation of $U_i$.
The outcome does not depend on the detailed order of the outcomes, 
but only on the number of 0s or 1s in the outcomes.
Counting the number of 1s as $s$, the final state takes two forms
\bea & \text{even} \; n: & \alpha \I + \frac{U(\psi)}{(d^2-1)^s}, \\ 
     & \text{odd} \; n:  & \beta  \I - \frac{U(\psi)}{(d^2-1)^s},
\eea 
for $\alpha$ and $\beta$ as normalization constants, 
and $U=\cdots U_2 U_1$ as the sequence of gates. 
That is, for any measurement outcome the final state
can be used to compute observable quantity from 
$|\bra \psi_\text{out}|\cdots U_2 U_1|\psi_\text{in}\ket|^2$,
hence showing the universality of OQT.  
Therefore, the OQT serves as the central protocol 
for program composition, as well as transmission.



There are a few novel features of OQT. 
First, there is a temporal order for a sequence of teleportation 
due to the byproduct correction, 
yet there is no temporal order for OQT. 
This brings a huge advantage of parallelism.
Second, teleportation can be used to reduce the circuit depth on a qubit,
and this also carries over to OQT. 
If the physical carrier (such as an atom) is acted upon by a long sequence of gates,
there can be heating issue or control problems. 
Teleporting the information to other fresh qubits can mitigate the issue.
While teleportation requires more storage in the form of ebits, 
but if qubits can be refreshed quickly after measurement, then they can be reused. 
That is to say, rapid qubit reset and reuse can minimize space overhead, 
making OQT an efficient space-time tradeoff for quantum computation.

We also see that the signal $U(\psi)$ comes with a factor $1/(d^2-1)^s$.
One would note that the state appears to be similar with the pseudo-pure state 
in the NMR platform~\cite{KL98}.
This is not a problem in the fault-tolerant setting where accuracy is guaranteed,
or when $s$ is relatively small.
However, if this is not the case extracting outcome from this signal could be hard.
One strategy to increase the stability of the outcome,
depending on the application tasks, 
is to use a hybrid of QT (including state and gate teleportation) and OQT. 
That is, only some gates are taken as black boxes in the simulated gate sequence
$\cdots U_2 U_1$.
The usage of QT will introduce a temporal order hence increase the circuit depth.
Therefore, it is an interesting task to design a teleportation network of QT and OQT 
depending on practical figure of merits. 


\section{Oblivious quantum control}
\label{sec:oqc}

In this section, we introduce oblivious quantum control (OQC),
which is to add quantum control to unknown quantum gates. 
This task is usually known as control over unknown gates~\cite{AFC14},
and here we coin the new name to highlight the role of being oblivious.
A no-control theorem was also found,
but different from the no-programming theorem, 
this no-go can be circumvented relatively easier~\cite{AFC14,W22_qvn},
and here we find a scheme to achieve it. 

\begin{figure}
    \centering
    \includegraphics[width=0.8\linewidth]{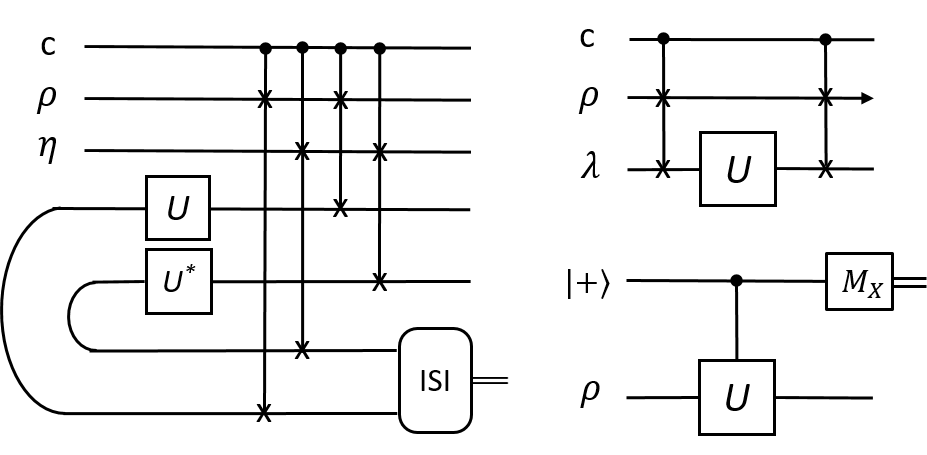}
    \caption{The oblivious quantum control scheme: our method (left)
    and the original method with $|\lambda\ket$ 
    as the `flag' (top right). 
    The DQC1 algorithm (bottom right) can be viewed as an application.}
    \label{fig:oqc}
\end{figure}

The task is to convert an arbitrary unknown unitary operation $U$ to its controlled version 
\be \wedge_U=P_0\otimes \I + P_1 \otimes U, \ee 
for $P_0$, $P_1$ as qubit projectors onto state $|0\ket$, $|1\ket$.
The reason for the no-control theorem~\cite{AFC14} 
is that the unphysical global phase of $U$ will be converted to a physical one in $\wedge_U$.
To fix this ambiguity, 
it was originally noted by Kitaev~\cite{KSV02} that 
if an eigenstate $|\lambda\ket$ and eigenvalue of $U$ is known,
serving as a `flag', 
then there is a way to realize this task.
See Fig.~\ref{fig:oqc}.
Such a `flag' condition appears simple but could be hard to realize, however.
For instance, finding an eigenstate and eigenvalue of a large $U=e^{itH}$ for a certain Hamiltonian $H$ 
is not an easy task.
In this section, we present a scheme that naturally satisfies this flag condition 
by extending the usage of $U$.

Our scheme is based on the following fact:
the ebit $|\omega\ket$ is an eigenstate of $U\otimes U^*$ with 
\be (U\otimes U^*) |\omega\ket = |\omega\ket. \ee 
For simplicity, let's first denote $\hat{U}:=U\otimes U^*$ and consider the task 
$\hat{U} \mapsto \wedge_{\hat{U}}$.
This can be achieved since the ebit $|\omega\ket$ serves as the flag. 
See Fig.~\ref{fig:oqc}(left).
The input for the data register is a product state $\rho \otimes \eta$ with $\eta$ as 
the ancillary state,
so that it is clear to see $U\mapsto \wedge_U$ is realized in the proper subspace
(for c and $\rho$).
The freedom of choosing $\eta$ can be used in quantum algorithms. 

In our setting, 
the gate $U$ can be given as a Choi state $|U\ket$.
The flag state $|\omega\ket$ can be realized via ISI scheme. 
Observing that 
\be \omega^{\perp}:= \frac{\I- |\omega\ket\bra \omega|}{d-1}\ee 
is also an eigenstate of $\hat{U}$,
so for both outcomes in the ISI scheme,
the flag condition is satisfied. 

Our scheme of oblivious quantum control (OQC) 
can be generalized to high-dimensional control to realize a so-called multiplexer
\be U=\sum_i P_i \otimes U_i,\ee  
for $P_i$ as qudit projectors.
A straightforward scheme is to decompose a multiplexer as a sequence of binary control as above,
and then realize each of them. 
We also see that both $U$ and $U^*$ are needed. 
This can be satisfied by providing them on the first hand,
or use the rebit approach as we have discussed in Sec.~\ref{sec:ope}.

\section{Oblivious quantum algorithms}
\label{sec:oqa}

In this section, we study a few oblivious quantum algorithms that
take OQT and OQC as primitives,
and these algorithms can be further used to construct more complicated 
algorithms or protocols. 

Quantum algorithms fundamentally operate by 
manipulating the amplitudes and phases of quantum states. 
Therefore, here we study the oblivious computation of state overlaps,
phase estimation, amplitude amplification, and generation of superposition. 
Among these, the oblivious amplitude amplification (OAA)~\cite{BCC+14}
has been well established, and 
it actually motivates our study of oblivious version of quantum algorithms.

\subsection{Oblivious DQC1}
\label{sec:odqc}

The original DQC1 algorithm~\cite{KL98},
i.e., deterministic quantum computing with one qubit,
relies on controlled operation to evaluate $\text{tr} U$ for a unitary operator $U$.
See Fig.~\ref{fig:oqc}, bottom right.
It is closely related to the Hadamard test 
and can also be used to compute overlap between states. 
When the input $\rho$ is a pure state $|\psi\ket$, 
the state before the Pauli $X$ measurement $M_X$ is 
\be |0\ket \frac{\I + U}{2} |\psi\ket + |1\ket \frac{\I - U}{2} |\psi\ket.  \ee 
The norm of $\frac{\I + U}{2} |\psi\ket$ is the probability $p_0$ for outcome 0, which is 
\be p_0=\frac{1}{2} (1+ \bra \psi| \text{Re} U |\psi\ket). \ee 
With the same circuit, but for the Pauli $Y$ measurement $M_Y$, 
it is easy to obtain the corresponding probability
\be p_0=\frac{1}{2}(1+ \bra \psi| \text{Im} U |\psi\ket). \ee 
From the probabilities above, the value of the overlap $\bra \psi| U |\psi\ket$ can be obtained.
Choosing the input $|\psi\ket$ as a mixed state $\rho$ will produce $\text{tr}(U\rho)$,
and the trace $\text{tr} U$ is obtained for the completely mixed state $\I/d$, in particular.
When $|\psi\ket$ is an eigenstate of $U$,
$\bra \psi| U |\psi\ket$ is the eigenvalue. 
This extends to Kitaev's quantum phase estimation (QPE)~\cite{NC00,KSV02},
and is used for Shor's algorithm~\cite{Sho94} and many others.

With the oblivious quantum control scheme,
we can obtain the oblivious DQC1 algorithm.
The unitary $U$ can be unknown or of a very large size.
The controlled module $\wedge_U$ is performed obliviously
and what is evaluated is the value of the product $\text{tr}(U\rho) \text{tr}(U^* \eta)$,
instead of $\text{tr}(U\rho)$.
The ancillary state $\eta$ can be chosen to eliminate a factor, e.g.,
by setting it as $|0\ket$ and the factor $\text{tr}(U^* \eta)$ becomes
$\bra 0| U^*|0\ket$, which could be easy to compute in practice. 
That is, $U$ is not completely unknown but the available information
is vanishingly limited. 

When both $|\psi\ket$ and $U$ are unknown, 
this algorithm can compute $\bra \psi| U |\psi\ket$.
Denote $|\phi\ket=U |\psi\ket$, this becomes $\bra \psi|\phi\ket$.
However, for two unknown states $|\psi\ket$ and $|\phi\ket$, 
there is no algorithm to compute $\bra \psi|\phi\ket$
since there is an ambiguity of the global phase. 
The ODQC1 algorithm avoid this since $U$ is not completely unknown,
and the relation $|\phi\ket=U |\psi\ket$ diminishes such an ambiguity.
A closely related algorithm is the SWAP test~\cite{BCW+01}, 
which can obliviously compute $|\bra \psi|\phi\ket|^2$, 
instead of $\bra \psi| U |\psi\ket$.
For the case of two Choi states $|U_1\ket$ and $|U_2\ket$, 
this becomes $|\text{tr}(U_1^\dagger U_2)|^2$.
Actually, we can use teleportation to achieve probabilistic generation of
$|U_1^\dagger U_2\ket$ from $|U_1\ket$ and $|U_2\ket$,
and the ODQC1 algorithm can compute the value $\text{tr}(U_1^\dagger U_2)$.


\subsection{Oblivious amplitude amplification}
\label{sec:oaa}

The oblivious amplitude amplification (OAA)~\cite{BCC+14} has been developed 
as the oblivious extension of amplitude amplification (AA)~\cite{BHM02},
and serves as an important primitive for oblivious quantum computing.
Combined with quantum phase estimation (QPE), 
it can also perform amplitude estimation.
Here we review its content,
and present a scheme that combines OAA with OQT to reduce the circuit depth.

In order to obtain the action of a unitary $U$ on data state $|\psi\ket$,
a unitary circuit $G$ acting on a control system and a data system with 
\be G|0\ket |\psi\ket = \sqrt{p}|0\ket U |\psi\ket + \sqrt{1-p} |\Phi' \ket \ee
is constructed, for a probability parameter $p\in (0,1)$
which is independent of $|\psi\ket$, and $\bra 0|\Phi' \ket=0$.
The goal is to prompt $p$ close to 1.
Treating $G$ as the dilation of a channel,
the operator $\sqrt{p} U$ is a Kraus operator,
and it is unnecessary to know other Kraus operators, i.e., 
the form of $|\Phi' \ket$.

\begin{figure}
    \centering
    \includegraphics[width=0.5\linewidth]{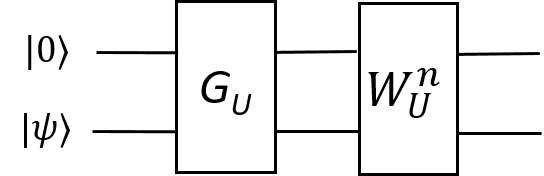}
    \caption{The circuit for amplitude amplification (AA) algorithm and also OAA algorithm.}
    \label{fig:oaa}
\end{figure}

The OAA constructed a walk operator $W=-GRG^\dagger R$ for $R$ acting 
on the control unit and $R=2\Pi-\I$ with $\Pi=P_0\otimes\I$.
For instance, $R$ is the Pauli $Z$ operator for a qubit controller. 
Let $\sqrt{p}=\sin\theta$ for $\theta\in(0,\frac{\pi}{2})$,
it proves that 
\be W^n G|0\ket |\psi\ket = \sin[(2n+1)\theta]|0\ket U |\psi\ket 
+\cos[(2n+1)\theta] |\Phi' \ket,  \ee 
and for $n\in O(1/\sqrt{p})$
the success rate can be boosted arbitrarily close to 1 to realize $U |\psi\ket $.
See Fig.~\ref{fig:oaa}.
In the case that $p$ is unknown on the first hand,
QPE and its oblivious version can be used to estimate $p$,
which can henceforth be amplified. 


It is also insightful to put OAA in the context of quantum programming (Eq.~(\ref{eq:cpu})).  
The program $U$ is not pre-stored as a state,
instead, it is stored in the circuit operations $G$ and $W$.
Therefore, it can generate $U|\psi\ket$ obliviously 
without violating the no-programming theorem~\cite{NC97}.

The OAA and also standard AA increase the circuit depth significantly 
by using iterative walk operators. 
This can be reduced to the original one (as $G$) 
by using OQT if we only need observable measurement on state $U|\psi\ket$, 
instead of the state itself.
The operators $G$, $GR$, and $G^\dagger R$ can be stored as Choi program states,
and OQT on them can compose them together. 
This needs $O(\sqrt{p})$ program states,
and this is quadratically better than an incoherent method, 
which measures the ancilla and repeat if a failure occurs.

Furthermore, the quantum singular-value transform (QSVT) algorithm 
has been developed as a generalization of AA~\cite{GSLW19}, 
which is the special case of only one singular value, the parameter $p$.
The QSVT algorithm converts a matrix $A$ with the singular-value transformation $A=W\Sigma V^\dagger$ 
to another matrix $B=W P(\Sigma) V^\dagger$ for $P(\Sigma)$ denoting a proper polynomial function of $\Sigma$.
The structure of QSVT is similar with AA, 
hence we can also use OQT to reduce its circuit depth by using stored programs
for the unitary $U$ that encodes $A$ and the reflection operators. 

\subsection{Oblivious quantum superposition}
\label{sec:oqs}

Combining the primitives above
and also the linear combination of unitary (LCU) operation algorithm~\cite{Long11,BCC+14,BCC+15},
we can realize oblivious quantum superposition (OQS) of unknown quantum states 
under proper conditions. 
This does not violate the no-go theorem~\cite{OGH16} since the states are not completely unknown.
We present two schemes with one relying on AA 
while the other relying on OAA,
while both of them rely on oblivious LCU that combines standard LCU and OQC.

The LCU can be viewed as an extension of the DQC1 circuit
with a multiple control of a set of gates $U_i$, i.e., a multiplexer.
By expressing an operator $C=\sum_i c_i U_i$ as a superposition of other unitary gates $U_i$,
the LCU circuit 
in general realizes $C$ for a success probability
\be p=\bra \psi|C^\dagger C|\psi\ket \ee 
and $C=\sum_i \alpha_i \beta_i U_i$, and $A|0\ket=\sum_i \alpha_i|i\ket$,
$B|i\ket=\sum_j \beta_{ij} |j\ket$, and $\beta_i :=\beta_{i0}$.
The coefficients can be chosen to satisfy $c_i=\alpha_i \beta_i$.
Given $U_i$ as oracles, the LCU circuit becomes oblivious. 
The walk operator in AA or OAA will also use $U_i^\dagger$,
and this can be resolved by the methods we presented for OQC in Sec.~\ref{sec:oqc}.

For the first scheme of oblivious superposition, 
the task is to generate a state $|\psi\ket \propto \sum_i c_i |\psi_i\ket$
with given known superposition coefficients $c_i$
while unknown set of states $\{|\psi_i\ket\}$, 
but $|\psi_i\ket=U_i|0\ket$ for a fixed state $|0\ket$.
As the state $|0\ket$ is known, 
the success probability can be boosted by AA algorithm. 

For the second scheme, the task is to apply an oblivious LCU circuit $C=\sum_i c_i U_i$
on an unknown state. 
In this case, we have to request $C$ being proportional to a unitary operator,
$C=\sqrt{p}U$, 
and then this fits into the framework of OAA,
which can be used to boost the success probability $p$ close to 1.

\section{Physical realizations}
\label{sec:phys}

In this section, we study the physical realizations of OQT, OQC, and others. 
All the tasks can be well described and understood as in the circuit model,
such as the creation of program states and measurement on them.
We assess the main cost of obliviousness by comparing to 
their non-oblivious counterparts of the algorithms above,
and layout the technical requirements on current quantum platforms.



\subsection{Cost of obliviousness}

There are common features for the primitive algorithms above, 
and we summarize them as follows:

\begin{itemize}
    \item The OPE, OQT schemes: they need an indirect binary projective measurement,
    which needs a Toffoli-type gate, also known as a multiple-control Toffoli gate. 
    \item The OQC, ODQC1, OQS schemes: they need the controlled-swap gate, also known as Fredkin gate,
    which can be further realized by Toffoli-type gates. 
    \item The OAA, OQS schemes: it needs a longer circuit which means more Fredkin gates.
    Using OQT to reduce the circuit depth would convert the cost to more ebits and also Toffoli-type gates.
\end{itemize}

Compared with an obvious preparation of a state $|\psi\ket$, 
the action of $U$, and the measurement of an observable on $U|\psi\ket$,
the central requirements for the oblivious version are the ebits and Toffoli-type gates,
besides ancillary qubits. 
The costs shall be clear from the details of the algorithms studied above, 
and below we further study the cost for realizing Toffoli-type gates.
Such a gate with $m$ control lines and a single target is often denoted as $\text{C}^m\text{NOT}$.

\begin{figure}
    \centering
    \includegraphics[width=0.7\linewidth]{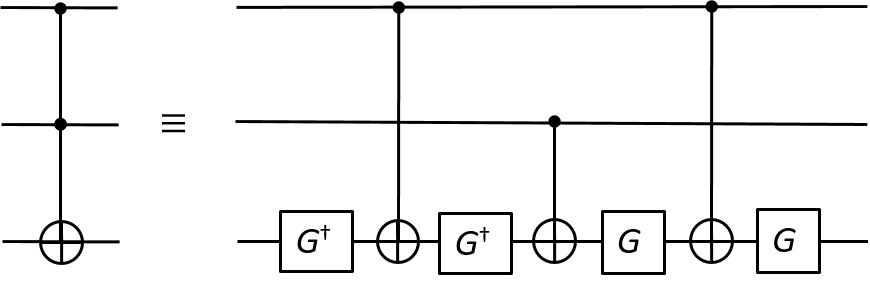}
    \caption{Circuit to realize the Toffoli gate with three CNOTs 
    when it is used for measurement.
    The gate $G$ is the rotation $R_Y(\frac{\pi}{4})$ around Y axis for $\frac{\pi}{4}$ angle.}
    \label{fig:tof}
\end{figure}

The standard Toffoli gate is a controlled CNOT gate $\text{C}^2\text{NOT}$,
with CNOT as the controlled Pauli $X$ gate. 
There are methods to directly realize Toffoli-type gates~\cite{LKS+19,KM20,KMN+22},
which will significantly reduce the circuit depth.
However, these techniques are still not mature yet. 
Instead, one can use gate compiling technique to decompose a Toffoli-type gate,
and in general, a multi-control gate~\cite{BBC+95,SM13}. 
Such circuits are efficient and using ancillary qubits 
can even reduce the circuit cost. 
The circuit cost gets smaller
if a Toffoli-type gate sits at the boundary of a circuit.
As shown in Fig.~\ref{fig:tof}, 
the Toffoli gate can be realized by qubit gates and three CNOTs~\cite{ICK+16},
instead of six,
if it is used for the ISI and OQT measurement schemes. 

\begin{figure}
    \centering
    \includegraphics[width=0.5\linewidth]{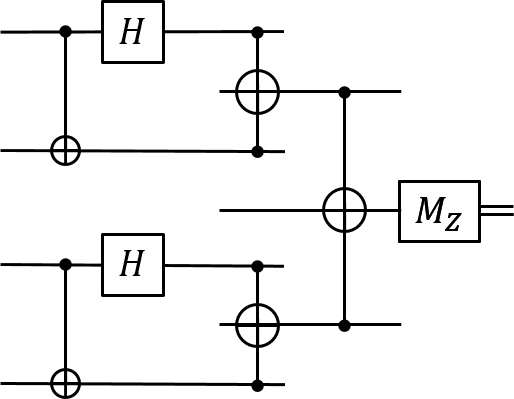}
    \caption{High-dimensional indirect binary Bell measurement for OQT.
    The circuit shown contains two local parts, 
    which can be extended to the multi-party cases.
    We do not specify the ancillary states for simplicity.}
    \label{fig:bm2}
\end{figure}

There is a systematic way to construct the OQT.
A high-dimensional OQT can be constructed from low-dimensional ones.
For instance, a two-partite OQT is shown in Fig.~\ref{fig:bm2}.
The total parity $k=i\cdot j$ is from the local parities $i$ and $j$,
and this construction easily extends to more general settings.
The cost is efficient with respect to the number of local parities,
i.e. the circuit width. 

Alternatively, this measurement procedure 
for OQT can be simplified by using more samples. 
For example, the two-partite OQT in Fig.~\ref{fig:bm2} 
merely realizes a measurement with projectors $P_{00}$ and $P_{\bar{00}}$,
for $P_{00}=P_{0}P_{0}$ and $P_{\bar{00}}$ as its complement,
while here 0 denotes a local parity information.
This can be realized by local measurements $\{P_{0},P_{\bar{0}}\}$ but with more runs 
to collect each measurement result, and finally gather the results 
as two sets for $P_{00}$ and $P_{\bar{00}}$, respectively.
It is easy to see, however, the number of samples grows exponentially with the number of local parts. 
So this scheme only works for relatively low-dimensional cases.
Also note that the two-qubit OQT can also be simulated by the usual two-qubit Bell measurement
(see Fig.~\ref{fig:tele}),
but only recording the Pauli byproduct $\sigma_i$ without correcting them.
This also requires more samples to gather the result for each byproduct $\sigma_i$. 

\subsection{Gates and qubits}

Here we study how to realize our schemes above in currently available platforms
for universal quantum computing.
We find that although being challenging, such as realizing Toffoli-type gates,
these operations indeed can be realized,
and there are notable differences across these platforms.
At present, there are a few leading quantum platforms with mainly two types of qubits:
\begin{itemize}
    \item Matter-qubit platform: such as superconducting processors, trapped ions, and cold atoms.
    The gates are generated on-demand while the qubits are hardware in a chip.
    \item Photon-qubit platform: linear optics.
    The gates are hardware in a chip while the qubits are generated on-demand from lasers.
\end{itemize}

For both of them states are mostly generated on-demand, 
since there is no self-correcting quantum memory yet to store qubits for long-enough time~\cite{BLP16}. 
Active quantum error-correction is required to extend the coherence time. 
A qubit can be either encoded in discrete degree of freedom or in continuous variable, 
while here we do not specify the details. 

For matter-qubit platforms, primitive gates such as qubit rotations and CNOT
can be realized deterministically. 
This can be used to prepare ebits and Choi program states.
To realize OQT, the low-dimensional case on two qubits 
could be simpler without using a Toffoli gate.
Such a parity measurement is to distinguish singlet from triplet,
which is not hard to perform. 
For instance, for superconducting qubits or spin qubits in quantum dots
the energy splitting $\hbar \omega$ of a qubit and the coupling strength $g$ between qubits 
can be tuned, 
so that an energy splitting $\Delta$ between singlet and triplet can be induced 
(See Ref.~\cite{SB25} and references therein).
The triplet states are nearly degenerate. 
For high-dimensional case and also OQC, 
Toffoli-type gates need to be realized which may have smaller values of fidelity. 

To realize teleportation, the interface between matter-qubit and photons 
may be needed for long-distance communication.
There are available techniques, for instance, 
states of superconducting qubits can be firstly translated to 
microwave resonators, and then use electro-optic transducer 
to convert into photonic states~\cite{JYL25,AYH25}.
For short-distance situations such as within a chip,
there are techniques to physically move qubits~\cite{BEH23,BEG24}.

For photon-qubit platforms, 
a scalable approach is the cluster-state quantum computing~\cite{RB01}
rather than the circuit model
due to the probabilistic nature of entangling gates~\cite{KLM01}.
It realizes a gate sequence by a sequence of gate teleportation. 
The smallest cluster is just an ebit.
For polarization or dual-rail encoded qubits,
ebits are relatively easy to prepare, 
e.g., generated via spontaneous parametric down-conversion
(SPDC) scheme~\cite{NC00} without using CNOT gates. 
Starting from ebits, 
a cluster state can be generated by the so-called fusion operations~\cite{BR05}.
For hardware, the region for generation of cluster state 
and the region for implementation of measurements can be separated.
These regions or chips can be connected by optical fibers~\cite{Abu24}, for instance. 
Photons can be transmitted directly or via teleportation,
which is a merit of the photon-qubit platform.

Another notable feature is that photonic gates are hardware.
This means that a gate $U$ can be stored as a hardware circuit, denoted as $H(U)$, 
instead of a Choi state $|U\ket$.
The hardware gate $H(U)$ makes it possible to obliviously
load it directly acting on input photons 
and realize $\wedge_U$ since there is a natural direct-sum structure 
of the Hilbert space~\cite{AFC14}.
However, this approach is not scalable due to the same reason for 
deterministically realizing entangling gates.
For instance, although beam splitters are universal to realize any SU($N$) operations,
the number of them scales exponentially with the number of qubits $n$ for $N=2^n$~\cite{RZB94}.
Therefore, in general we need to employ the cluster-state approach,
and a quantum program $|U\ket$ can be generated by measurements on a cluster state.
This requires a massive amount of classical side processing of measurement feedforward,
and also measurement devices.

\begin{figure}[t!]
    \centering
    \includegraphics[width=0.8\linewidth]{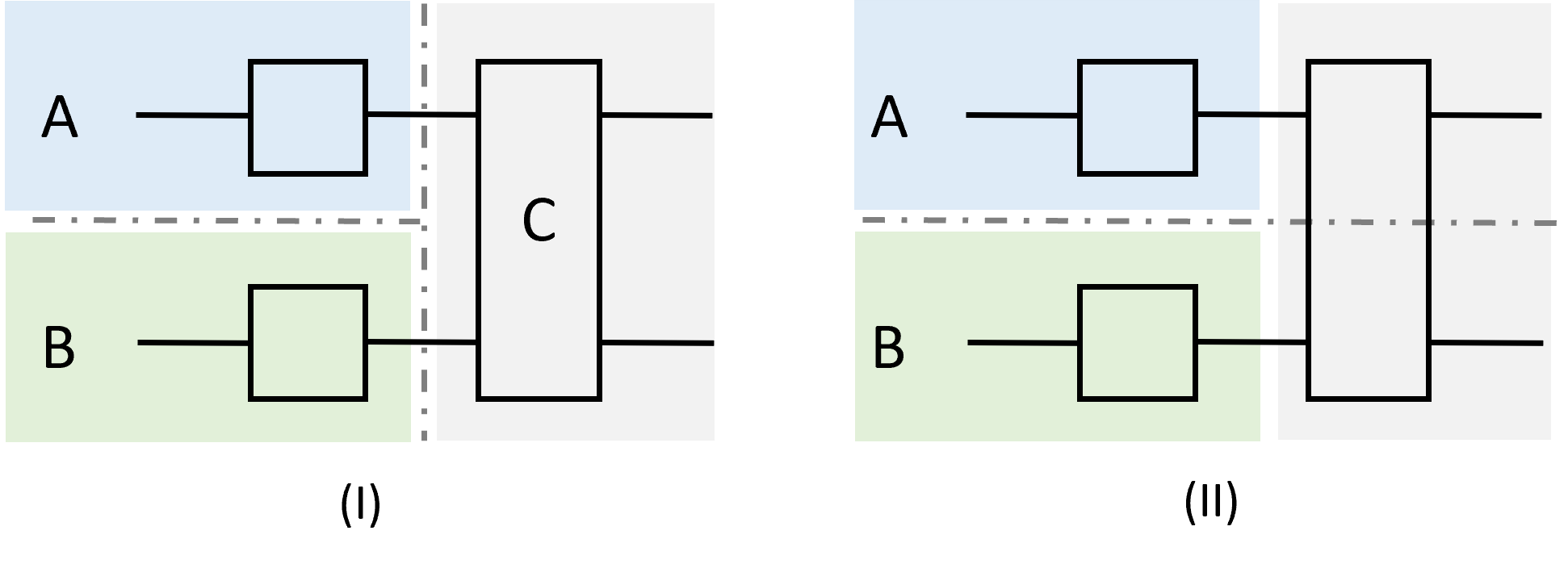}
    \caption{The schematic for a distributed quantum computing with three participants.
    The dashed lines set the locality.}
    \label{fig:dqc}
\end{figure}

The realization of photonic OQT also needs a cluster state 
to simulate the Toffoli-type gates in it.
Beyond this general approach, 
we find that there could be easier schemes. 
The primary fusion measurement divides into 
two flavors:
the type-I fusion which is a 1-bit teleportation~\cite{ZLC00},
and the type-II fusion which is the usual 2-bit teleportation.
They both fail with $50\%$ percent of probability.
It is also known that the type-II fusion is from the 
Hong-Ou-Mandel interference effect~\cite{HOM87}.
When the type-II fusion is used to
distinguish the anti-symmetric singlet from 
the symmetric triplet, 
without making further observations among the triplet states,
it actually realizes the OQT on two qubits without using a Toffoli gate.
Therefore, it remains to see if the high-dimensional OQT
can be efficiently simulated by multi-photon interference using
a network of beam splitters, 
also known as Bell multiport beam splitter~\cite{ZZH97,TTM10}, 
whose design is highly nontrivial. 

\section{Distributed quantum computation}
\label{sec:dqc}

In this section, we apply the oblivious quantum algorithms in the setting of distributed 
quantum computing. 
We first study a primary protocol and then generalize it,
and discuss more features of our protocol.


\subsection{Primary protocol}

\begin{figure}[t!]
    \centering
    \includegraphics[width=0.8\linewidth]{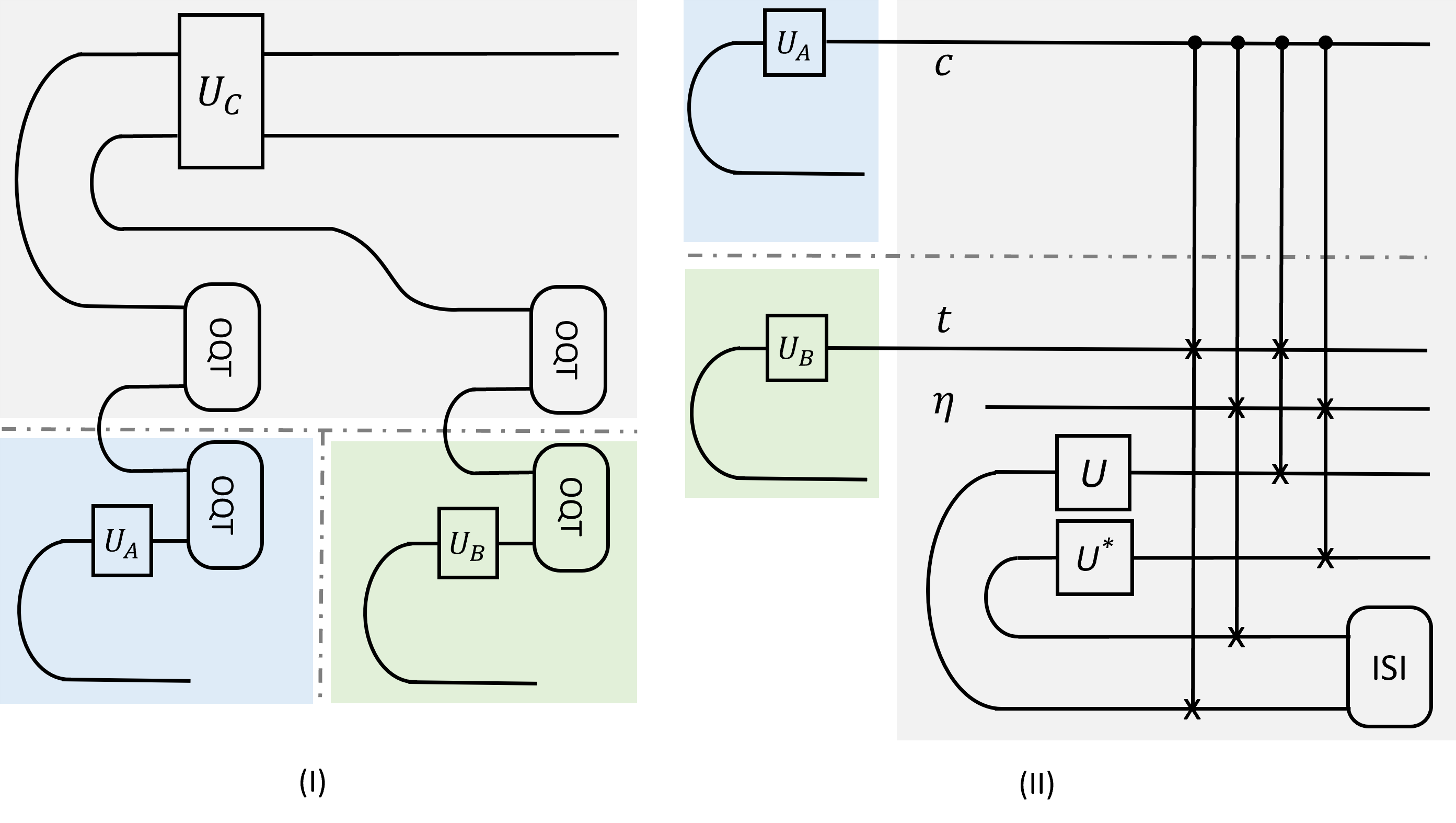}
    \caption{The schematic for the realization with OQT or OQC for
    the distributed quantum computing with three participants in Fig.~\ref{fig:dqc}.
    Note we only show one $\wedge_U$ for the OQC scheme (II) for simplicity. 
    The dashed lines set the locality.
    }
    \label{fig:dqc2}
\end{figure}

We first consider the two-party and tri-party cases, 
and then generalize to the multi-party setting. 
The protocol is illustrated in Fig.~\ref{fig:dqc} and Fig.~\ref{fig:dqc2}.
The two-party case is a reduction of it.
Suppose Alice and Bob aims to compute some observable value $O$ on a state 
$U|\psi\ket$, but now $U=U_B U_A$ for $U_A$ ($U_B$) held by Alice (Bob) 
in the form of Choi state $|U_A\ket$ ($|U_B\ket$).
The initial state $|\psi\ket$ is prepared by Alice,
while the measurement for observable $O$ is performed by Bob.
In order to gather statistics, multiple runs are needed.
The protocol looks symmetric with respect to Alice and Bob,
but it is not due to the input-output separation. 

It uses communicating ebits to perform OQT. 
This is due to a consideration of hardware realization
since it may be hard to directly transmit some part of the stored programs 
in the memory unit,
hence two OQT actions are needed.
Physically, the OQT may involve interactions between flying qubits such as photons 
and solid-state qubits, e.g., Refs.~\cite{JYL25,AYH25}.

For the tri-party case, 
the dimensions of the composed programs could be different.
We find there are two schemes:
\begin{itemize}
    \item I): The nonlocal part is a separate station C, as in Fig.~\ref{fig:dqc}, panel I;
    \item II): The nonlocal part is a nonlocal operation on A and B, 
            as in Fig.~\ref{fig:dqc}, panel II.
\end{itemize}
Their realizations are shown in Fig.~\ref{fig:dqc2}.
For scheme I,
a concatenation of the parity measurements is needed, namely, 
the parity $i$ for the OQT between $|U_A\ket$ and $|U_C\ket$ 
and the parity $j$ for the OQT between $|U_B\ket$ and $|U_C\ket$ 
is further summed to a total parity $k$.
This serves as the OQT between the program $|U_C\ket$ and the combined  
program $|U_A\ket|U_B\ket$.
The construction of such high-dimensional OQT has been discussed in Sec.~\ref{sec:phys}.

For scheme II, the nonlocal gate could be a product of controlled gates
according to general gate compilation~\cite{BBC+95,SM13}.
For instance, if it is a $\wedge_U$ for A as the control $c$ and B as the target $t$
and B holds the program $|U\ket$,
then we can apply a remote OQC by using remote Toffoli-type gates. 
This finally reduces to remote CNOT gates,
which can be realized by gate teleportation~\cite{GC99}.
Each CNOT gate consumes one ebit. 
More importantly, this scheme avoids the OQT, despite the increase of circuit depth 
due to the byproduct correction in gate teleportation.
This is a case that we discussed in Sec.~\ref{sec:oqt} for constructing 
a hybrid QT and OQT network.

\subsection{General protocol}

\begin{figure}[t!]
    \centering
    \includegraphics[width=0.7\linewidth]{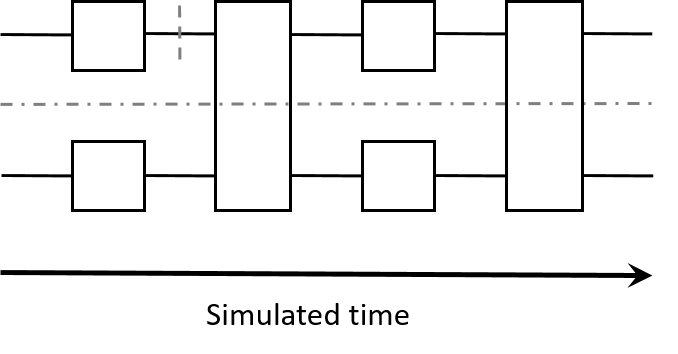}
    \caption{The schematic for a multi-party distributed oblivious quantum computing.
    Each box could be a separable participant.
    The dashed lines are only for illustration of locality.
    In general, this is used in hybrid classical-quantum distributed algorithm.
    }
    \label{fig:dqc3}
\end{figure}

The protocol above can be extended to the multi-party setting,
as illustrated in Fig.~\ref{fig:dqc3}.
We now describe the general algorithmic structure of it. 
When there are quantum programs, i.e., 
Choi states $\{\omega_{\C E_i}\}$ for channels $\{\C E_i\}$ as input, 
the quantum circuit acting on them is generally known as 
quantum superchannel or comb~\cite{CDP08a}. 
Such circuits can be viewed as quantum programming or 
quantum super-algorithms~\cite{W22_qvn,LWLW23}, although they can be 
described as usual quantum circuits. 
On top of the quantum circuit,  
there can be a classical algorithm $\C A$ which 
is used to optimize the parameters $\vec{\theta}$
in a quantum circuit $U(\vec{\theta})$.
The algorithm $\C A$ is a function $f(\vec{x})$ of the measurement outcome 
$\vec{x}=\text{tr}(O\rho_{\vec{\theta}})$ from the circuit~\cite{W24rev}.
The OQC and other oblivious quantum algorithms,
as well as explicit operations, are a part of the superchannel,
and the OQT can be used to break the whole high-depth circuit 
into smaller blocks for each program $\omega_{\C E_i}$, for instance. 
As the OQT guarantees the final outcome $\vec{x}$,
not necessarily the final state,
this naturally fits into the framework of hybrid classical-quantum algorithm.

As mentioned in Sec.~\ref{sec:intro},
our protocol suits many practical settings. 
The programs $\omega_{\C E_i}$ can be the encoding of classical or quantum data
depending on the problem to be solved.
Examples include quantum channel estimation, 
quantum machine learning and optimization, and secure quantum computing.  
The unknown objects can be the unknown channel to be estimated, 
some pre-stored data or subroutines, or private input from users, respectively. 

For instance, it is known that for quantum channel estimation or discrimination
the sequential scheme by superchannel is necessary for non-unitary channels~\cite{CAP08}.
Using our scheme, this breaks down to three steps: 
firstly sending an ebit each to the unknown channel $\C E_i$ 
to generate a collection of $\omega_{\C E_i}$,
and apply a part of superchannel to each $\omega_{\C E_i}$,
and then use OQT to connect them.
All the steps are transversal, i.e., can be done in parallel. 
This greatly reduces the circuit depth while consuming more ebits and qubits.

\begin{figure}[t!]
    \centering
    \includegraphics[width=0.7\linewidth]{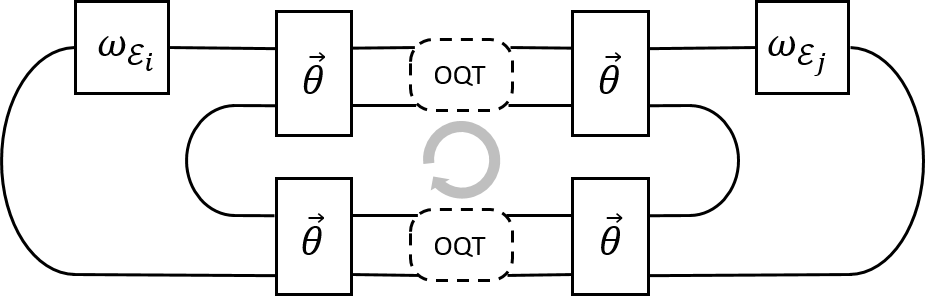}
    \caption{A schematic to illustrate the space-time tradeoff in OQT. 
    A parallel OQT can be realized by iterative OQT between two blocks if data and programs can be loaded quickly.
    Boxes with $\vec{\theta}$ and ebits represent operations in the quantum superchannel. 
    }
    \label{fig:combalg}
\end{figure}

As explained in Sec.~\ref{sec:oqt},
a space-time tradeoff applies for the usage of OQT if qubits can be refreshed. 
It is not hard to see the minimal number of samples is two,
shown in Fig.~\ref{fig:combalg},
which teleports the information back and forth. 
Now the OQT applies sequentially. 
A qubit needs to survive longer than the period of 
an OQT operation followed by the preparation of a program.


\subsection{Comparison with other protocols}
\label{sec:ob}

To better understand the features of our protocol and find more application settings,
in this section we compare our protocol with a few other relevant quantum computing protocols. 
This is briefly summarized in Table~\ref{tab:comp} and analyzed in details below. 
These protocols have close connection with quantum cryptography and secure quantum computing~\cite{KL20}.



First, let us recall our protocol in the bipartite case. 
The protocol, abbreviated as distributed black-box quantum computing (DBQC) in the Table,
compute overlaps $|\bra \psi_\text{o}|U_\text{o}U_\text{in}|\psi_\text{in} \ket|^2$ with a user Alice holding 
the input state $|\psi_\text{in}\ket$, the classical description of it $[\psi_\text{in}]$, 
and the program state $|U_\text{in}\ket$,
and a server Bob holding the final measurement specified by a set of $\{|\psi_\text{o}\ket\}$,
their classical description $[\psi_\text{o}]$, and another program state $|U_\text{o}\ket$.

This is directly distinct from the `obvious' protocol when 
Alice knows $[U_\text{in}]$ and Bob knows $[U_\text{o}]$. 
This is the distributed quantum computing (DQC) in the Table.
Actually, 
each participant can apply a gate $U_i$ directly 
and use QT to send its system to others. 
A nonlocal gate can be done remotely or belongs to a nonlocal participant. 
This setting is common and a few methods have been developed, 
such as distributed schemes for some quantum algorithms~\cite{ACR21,QLX24,TLW23,MGD23,Hwang_2025}
and the circuit knitting technique~\cite{PS23,UPR23,LMH23}. 
The obvious setting is easier due to the lack of obliviousness, 
and some techniques could also benefit DBQC. 
For instance, the circuit knitting aims to 
break down a nonlocal gate into a sum of local ones. 
An example is the expansion of a two-body gate in terms of Pauli gates 
\be U=\sum_{ij}u_{ij}\sigma_i \otimes \sigma_j, \ee 
as Pauli gates form an operator basis. 
The coefficients $u_{ij}$ are complex and understood as quasi-probabilities.
This will increase the sampling cost and it is easy to see 
the cost increases exponentially with the number of nonlocal gates. 
Therefore, similar with the local reduction of OQT in Sec.~\ref{sec:phys},
these techniques only work for relatively small circuits. 

In our protocol of DBQC, there is a weak sense of obliviousness,
so we need to further clarify this. 
In oblivious transfer (OT)~\cite{KL20} and OT-based quantum computing~\cite{Hay22}, 
a server (Bob) transmits two bit strings to a user (Alice), 
who can retrieve only one of them while Bob remains unaware of Alice’s choice.
The server can do a computation $U$ without knowing the input $|\psi_\text{in}\ket$ and output $|\psi_\text{o}\ket$.
The security of OT is complicated,
and we defer this to the Appendix~\ref{sec:bcot}.


In DBQC, a type of oblivious transfer of quantum information can be achieved.
Alice hides the state $U_A|\psi\ket$ separately as the program $|U_A\ket$ 
and the ISI measurement of $|\psi\ket$.
Bob hides his information in $|U_B\ket$. 
Alice and Bob will broadcast their outcomes of OQT,
and Bob also needs to send the measurement outcomes to Alice. 
Alice gets the final result. 
There is no way for Alice to know the whole information of $U_B$,
and neither for Bob to know the whole information of $U_A|\psi\ket$
due to the no-cloning theorem. 
The security holds for the passive adversary setting whereas 
the participants are honest but curious.
However, it does not hold when participants are malicious, e.g.,
Bob can send wrong measurement outcomes to Alice to spoil the task. 

In blind quantum computing (BQC)~\cite{BFK09}, 
a user knows the computing task while only calls a server as an oracle to 
execute the circuit $U$.
The server is blind or oblivious of the computing task. 
The final protocol we consider is a task in quantum von Neumann architecture (QvN)~\cite{W22_qvn},
where a server can generate a program $U$ and send it to a user as $|U\ket$. 
Given a limited amount of samples, 
the user can verify the program $U$ without learning it~\cite{MSG+22,LWLW23}, 
hence the user is oblivious.
Actually, the DBQC can be viewed as a distributed version of QvN,
with a circuit $U$ being split into a few parts held by separate participants.



\begin{table}[t!]
\footnotesize
    \centering
    \begin{tabular}{c|l|l} \Xhline{0.8pt}
              & user & server \\ \hline 
OT & $|\psi_\text{in}\ket$, $[\psi_\text{in}]$, $|\psi_\text{o}\ket$, $[\psi_\text{o}]$           
& $U$, $[U]$ \\ \hline 
BQC     & $|\psi_\text{in}\ket$, $[\psi_\text{in}]$, $|\psi_\text{o}\ket$, $[\psi_\text{o}]$, $[U]$    
& $U$   \\ \hline 
QvN       & $|\psi_\text{in}\ket$, $[\psi_\text{in}]$, $|\psi_\text{o}\ket$, $[\psi_\text{o}]$, 
$|U\ket$ & $U$, $[U]$ \\ \hline 
DBQC      & $|\psi_\text{in}\ket$, $[\psi_\text{in}]$, $|U_\text{in}\ket$                 & $|\psi_\text{o}\ket$, $[\psi_\text{o}]$, $|U_\text{o}\ket$ \\ \hline 
DQC       & $|\psi_\text{in}\ket$, $[\psi_\text{in}]$, $|U_\text{in}\ket$, $[U_\text{in}]$     & $|\psi_\text{o}\ket$, $[\psi_\text{o}]$, $|U_\text{o}\ket$, $[U_\text{o}]$ \\ \Xhline{0.8pt}
    \end{tabular}
    \caption{The comparison of a few quantum computing protocols for the bipartite setting.
    The abbreviations are OT (oblivious transfer), BQC (blind quantum computing), 
    QvN (quantum von Neumann architecture), DBQC (distributed black-box quantum computing),
    and DQC (distributed quantum computing).
    The details are found in the main text.
    }
    \label{tab:comp}
\end{table}

\section{Conclusion}


In this work, we proposed a protocol of universal distributed quantum computing with black-box subroutines.
The protocol is based on a few oblivious quantum algorithms that involve 
black-box subroutines.
We also analyzed the requirements of obliviousness and pointed out 
the potential application in various settings.



We also would like to address the differences and connections among a few 
computational notions. 
The very basic notion of universality, 
in a narrow sense, refers to the ability to realize an arbitrary $U$ within an accuracy $\epsilon$
by a process $C$, with the cost of $C$ being efficient 
with respect to $\epsilon$ and a proper measure of $U$~\cite{NC00}.
It does not prescribe the physical implementation mechanism for gate operations, 
nor does it inherently require programmability - the ability to 
store and subsequently execute gate operations as callable subroutines. 
Computation with unknown subroutines is oblivious,
and it can further lead to security in a distributed multi-party setting~\cite{CGS02}. 

The computational notions above are closely tied with 
quantum von Neumann architecture,
which includes modular hardware units for quantum CPU, control, memory/storage, 
communication, input and output. 
Realizing a true quantum von Neumann architecture 
will depend heavily on future hardware developments, 
particularly the emergence of specialized quantum memory systems~\cite{LWLW23}.
From another perspective, 
the current model of quantum computing is a form of ``in-memory" computing~\cite{SLK20} 
where the traditional distinction between processing and memory units disappears. 
The von Neumann architecture can also be understood and employed from a software perspective,
which refers to quantum programming with functional subroutines 
for the benefit of protection of programs and also higher-level integration.
In this regard,
our distributed protocol in this work can be viewed as a simulation of 
quantum von Neumann architecture.
Overall, our findings not only establish meaningful connections with 
existing distributed quantum computing protocols 
but also open new avenues for advancing both theoretical and experimental research in the field.

\section{Acknowledgement}
This work has been funded by
the National Natural Science Foundation of China under Grants
12447101, 12105343, and 62471368,
the Natural Science Foundation of Guangdong Province (Grant No. 2023A1515010671),  
and the Shaanxi Provincial Young Innovative Teams in Higher Education.
Suggestions from W.-Z. Cai, M. Hayashi, and Y.-D. Wu are greatly acknowledged.

\appendix 



\section{Extension to quantum channels and superchannels}
\label{sec:superc}

In the main text, we have presented a few schemes for the unitary case.
Here we extend them to the non-unitary channels. 
The channel-state duality~\cite{Cho75,Jam72} states that a channel 
$\C E$ can be represented as a Choi state 
\be \omega_{\C E} := \C E \otimes \I (\omega), \ee 
for $\omega:=|\omega\ket\bra \omega|$ as the Bell state.
From dilation a channel $\C E$ can also be realized by a unitary circuit $U$
requiring an additional ancilla initialized at $|0\ket$.



The general operations on Choi states are superchannels~\cite{CDP08a,WW23}.
The unitary dilation of a superchannel $\hat{\C S}$ requires two unitary 
operators $U_1$ and $U_2$ such that the Kraus operator-sum form is
\be \hat{\C S}(\omega_{\C E})=\sum_\mu S_\mu \omega_{\C E} S_\mu^\dagger\ee
with Kraus operators
\be S_\mu= \sum_m K_{2,m\mu} \otimes K_{1,m} \label{eq:s}\ee 
and $\sum_\mu S_\mu^\dagger S_\mu=\I$, 
$K_{1,m}=\bra m|U_1|0\ket$ and $K_{2,m\mu}=\bra m|U_2|\mu\ket$ 
are also Kraus operators. 
The sum over $m$ signifies the quantum memory between $U_1$ and $U_2$.
From an algorithmic point of view, 
the $U_1$ and $U_2$ form parametric quantum circuits that can be optimized. 

The oblivious quantum teleportation (OQT) easily extends to the non-unitary case. 
Two programs $\omega_{\C E_1}$ and $\omega_{\C E_2}$ can be composed 
to yield $\omega_{\C E_2 \C E_1}$ to guarantee observable measurement.
For the oblivious quantum control (OQC),
the situation is different.
If the dilated circuit $U$ for a channel $\C E$ is available,
we can use $U$ and OQC to generate $\wedge_U$,
and finally trace out the ancilla.
A superposition of channels $\{\C E_i\}$ can be realized as 
a superposition $\sum_i c_i U_i$ for a collection of $U_i$, 
and tracing out the ancilla leads to 
\be \sum_{ij} c_i c_j^* \text{tr}_a \left(U_i (\rho \otimes \rho_a) U_j^\dagger\right) \ee 
for $\rho_a:=|00\cdots 0\ket$ as the initial ancillary states,
and the trace $\text{tr}_a$ is over the ancilla.
It contains the `diagonal' terms $\sum_i |c_i|^2 \C E_i(\rho)$
and the interference terms for $i\neq j$.
The non-unitary OQC also applies to DQC1
which can evaluate 
tr($U(\rho\otimes |0\ket\bra 0|)$), which is tr($K_0\rho$)
for the first Kraus operator $K_0$ in $U$.
This extends DQC1 from evaluating the trace tr$U$ of unitary 
matrix to more general matrices.

\section{ Oblivious transfer and Bit commitment}
\label{sec:bcot}

In our scheme of distributed quantum computing, 
there is a sense of oblivious transfer,
but it is different from the traditional one. 
Therefore, we study this issue in more details. 

In classical cryptography, bit commitment (BC) and oblivious transfer (OT)
are two primary building blocks for more advanced protocols~\cite{KL20}. 
In BC, Alice commits a bit $b$ to Bob who cannot know its value,
and Alice will later on reveal its value without being able to change it.
In OT, Alice has two bits $b_0$ and $b_1$ while Bob needs one of them, $b_c$, with $c$ as his bit.
Bob only gets the requested bit without knowing the other one,
while Alice does not know his choice $c$.
Note here, for convenience, 
our assignment of Alice and Bob is different from that in the main text.

It is known that OT is universal for secure two-party classical computation,
and OT implies BC.
For quantum cryptography that uses qubits to encode bits,
the situation is different. 
It is known that 
quantum OT is equivalent to quantum BC,
and they are information-theoretically secure in the so-called noisy storage model~\cite{WST08},
otherwise they are not~\cite{May97,LC97}.  

To put our study in the more broader context of cryptography,
we first classify four categories of protocols:
\begin{enumerate}
    \item Transmit bits with classical means; this is the standard classical cryptography;
\item  Transmit bits with quantum means; this is the quantum cryptography;
\item  Transmit qubits with entanglement-assisted means; 
this refers to quantum cryptography of quantum information but with possible entanglement or 
classical correlations as assistance, for instance, 
the commitment of qubits, or known as qubit commitment~\cite{MPS18};
\item  Transmit qubits with purely quantum means. 
\end{enumerate}

From the perspective of channel capacity~\cite{Wat18},
with no surprise, the protocols above correspond to the classical capacity of classical channel, 
the private (or secure) capacity of quantum channel, 
the entanglement-assisted quantum capacity of quantum channel, 
and quantum capacity of quantum channel, respectively, 
ignoring the details of assisted classical communication. 

For quantum BC, it is well known a dishonest Alice can use entanglement to cheat:
Alice use bipartite non-orthogonal states $\psi_0$ and $\psi_1$ to encode 0 and 1, respectively, 
and send one local part to Bob,
and Bob cannot learn the committed bit from this local part,
but Alice can do local operations to switch between $\psi_0$ and $\psi_1$,
hence cheating. 
This requires Alice to have the ability to store qubits,
which is avoided in the noisy storage model~\cite{WST08}. 
The same cheating strategy carries over to the commitment of qubits~\cite{MPS18},
which we recall here. 
Alice aims to commit a qubit state $|\psi_\theta\ket=\frac{1}{\sqrt{2}}(|0\ket+e^{i\theta}|1\ket)$,
and first uses the Bell circuit to mask it in a Choi state 
$|\Psi_\theta\ket=\frac{1}{\sqrt{2}}(|00\ket+e^{i\theta}|11\ket)$.
Alice sends Bob half of the state which is completely mixed.
Bob's goal, say, is to check if Alice has made the correct guess of $\theta$,
but Alice still can freely change $\theta$ without being noticed by Bob. 
This cheating also applies to oblivious transfer of qubits. 

It is well known that quantum capacity is private against eavesdroppers, 
while it often assumes both the sender Alice and the receiver Bob are honest.
In a sense, the standard quantum communication is both committing and oblivious.
Namely, for commitment, Alice commits and sends a state $|\psi\ket$ directly.
There is no way for Alice to change it anymore,
neither for Bob to decode it until Alice tell him what it is, 
i.e., a classical description $[\psi]$ of it. 
For oblivious transfer, Alice and Bob first agreed upon an assignment $i\mapsto |\psi_i\ket$, 
for a set of index $i$ that does not reveal $\psi_i$,
and Bob only knows his index while Alice holds all the states without knowing them.
Bob tells Alice the index $i$ and Alice then sends the state $|\psi_i\ket$ to Bob.
Although Alice knows $i$, she would not know the state $|\psi_i\ket$.
For both protocols, 
Bob can finally use $[\psi]$ and quantum verification schemes to verify the state $|\psi\ket$.  
Our scheme for DBQC belongs to category (4),
which does not allow entanglement assistance, hence the entanglement attack does not apply.
However, it cannot tolerant dishonest participants who can use the wrong programs 
or send wrong measurement outcomes.
Whether our protocol can be boosted towards 
secure multi-party quantum computation remains an interesting task.




\end{spacing}
\bibliography{ext}{}

\begin{thebibliography}{10}
\expandafter\ifx\csname url\endcsname\relax
  \def\url#1{\texttt{#1}}\fi
\expandafter\ifx\csname urlprefix\endcsname\relax\def\urlprefix{URL }\fi
\expandafter\ifx\csname href\endcsname\relax
  \def\href#1#2{#2} \def\path#1{#1}\fi

\bibitem{LJL+10}
T.~D. Ladd, F.~Jelezko, R.~Laflamme, Y.~Nakamura, C.~Monroe, J.~L. O’Brien, Quantum computers, Nature 464~(7285) (2010) 45--53.

\bibitem{NC00}
M.~A. Nielsen, I.~L. Chuang, Quantum Computation and Quantum Information, Cambridge University Press, Cambridge U.K., 2000.

\bibitem{W24rev}
D.-S. Wang, Universal quantum computing models: a perspective of resource theory, Acta Phys. Sin. 73 (2024) 220302.

\bibitem{CAF24}
M.~Caleffi, M.~Amoretti, D.~Ferrari, et~al., Distributed quantum computing: a survey, Computer Networks 254 (2024) 110672.

\bibitem{BCD25}
D.~Barral, F.~J. Cardama, G.~Díaz-Camacho, et~al., Review of distributed quantum computing: From single {QPU} to high performance quantum computing, Computer Science Review 57 (2025) 100747.

\bibitem{BFK09}
A.~Broadbent, J.~Fitzsimons, E.~Kashefi, Universal blind quantum computation, in: in Proceedings of the 50th Annual Symposium on Foundations of Computer Science (IEEE Computer Society, Los Alamitos, CA, 2009), 2009, pp. 517--527.

\bibitem{NC97}
M.~A. Nielsen, I.~L. Chuang, Programmable quantum gate arrays, Phys. Rev. Lett. 79 (1997) 321--324.

\bibitem{W20_choi}
D.-S. Wang, Choi states, symmetry-based quantum gate teleportation, and stored-program quantum computing, Phys. Rev. A 101 (2020) 052311.

\bibitem{YRC20}
Y.~Yang, R.~Renner, G.~Chiribella, Optimal universal programming of unitary gates, Phys. Rev. Lett. 125 (2020) 210501.

\bibitem{W24_qvn}
D.-S. Wang, {A family of quantum von Neumann architecture}, Chin. Phys. B 33 (2024) 080302.

\bibitem{ACR21}
J.~Avron, O.~Casper, I.~Rozen, Quantum advantage and noise reduction in distributed quantum computing, Phys. Rev. A 104 (2021) 052404.

\bibitem{QLX24}
D.~Qiu, L.~Luo, L.~Xiao, {Distributed Grover's algorithm}, Theoretical Computer Science 993 (2024) 114461.

\bibitem{TLW23}
H.~Tang, B.~Li, G.~Wang, H.~Xu, C.~Li, A.~Barr, P.~Cappellaro, J.~Li, Communication-efficient quantum algorithm for distributed machine learning, Phys. Rev. Lett. 130 (2023) 150602.

\bibitem{MGD23}
S.~C. Marshall, C.~Gyurik, V.~Dunjko, High dimensional quantum machine learning with small quantum computers, Quantum 7 (2023) 1078.

\bibitem{Hwang_2025}
K.~Hwang, H.-T. Lim, Y.-S. Kim, D.~K. Park, Y.~Kim, Distributed quantum machine learning via classical communication, Quantum Science and Technology 10~(1) (2024) 015059.

\bibitem{PS23}
C.~Piveteau, D.~Sutter, Circuit knitting with classical communication, IEEE Trans. Inform. Theory 70 (2023) 3310797.

\bibitem{UPR23}
C.~Ufrecht, M.~Periyasamy, S.~Rietsch, D.~D. Scherer, A.~Plinge, C.~Mutschler, Cutting multi-control quantum gates with zx calculus, Quantum 7 (2023) 1147.

\bibitem{LMH23}
A.~Lowe, M.~Medvidović, A.~Hayes, et~al., Fast quantum circuit cutting with randomized measurements, Quantum 7 (2023) 934.

\bibitem{JYL25}
J.~Qiu, Y.~Liu, L.~Hu, et. al, Deterministic quantum state and gate teleportation between distant superconducting chips, Science Bulletin 70~(3) (2025) 351--358.

\bibitem{AYH25}
A.~Almanakly, B.~Yankelevich, M.~Hays, et~al., Deterministic remote entanglement using a chiral quantum interconnect, Nat. Phys.Https://doi.org/10.1038/s41567-025-02811-1 (2025).

\bibitem{GPA24}
A.~M. Gomez, T.~L. Patti, A.~Anandkumar, S.~F. Yelin, Near-term distributed quantum computation using mean-field corrections and auxiliary qubits, Quantum Sci. Technol. 9 (2024) 035022.

\bibitem{AFM24}
P.~Andres-Martinez, T.~Forrer, D.~Mills, et~al., Distributing circuits over heterogeneous, modular quantum computing network architectures, Quantum Sci. Technol. 9 (2024) 045021.

\bibitem{LZFD24}
T.-Y. Luo, Y.-Z. Zheng, X.~Fu, Y.-X. Deng, Automatic architecture design for distributed quantum computing, Chinese Phys. B 33 (2024) 120302.

\bibitem{Hay17}
M.~Hayashi, Quantum Information Theory: Mathematical Foundation, 2nd edition, Springer, 2017.

\bibitem{HH13}
D.~M. Harris, S.~L. Harris, Digital design and computer architecture, Elsevier, 2013.

\bibitem{KL20}
J.~Katz, Y.~Lindell (Eds.), Introduction to Modern Cryptography, Chapman and Hall/CRC (New York), 2020.

\bibitem{BCC+14}
D.~W. Berry, A.~M. Childs, R.~Cleve, R.~Kothari, R.~D. Somma, Exponential improvement in precision for simulating sparse hamiltonians, in: Proc. 46th ACM Symposium on Theory of Computing, 2014, p. 283.

\bibitem{SWP+17}
D.~T. Stephen, D.-S. Wang, A.~Prakash, T.-C. Wei, R.~Raussendorf, Computational power of symmetry-protected topological phases, Phys. Rev. Lett. 119 (2017) 010504.

\bibitem{WZ82}
W.~K. Wootters, W.~H. Zurek, A single quantum cannot be cloned, Nature 299 (1982) 802–803.

\bibitem{Die82}
D.~Dieks, Communication by {EPR} devices, Phys. Lett. A 92 (1982) 271.

\bibitem{May97}
D.~Mayers, Unconditionally secure quantum bit commitment is impossible, Phys. Rev. Lett. 78 (1997) 3414--3417.

\bibitem{LC97}
H.-K. Lo, H.~F. Chau, Is quantum bit commitment really possible?, Phys. Rev. Lett. 78 (1997) 3410--3413.

\bibitem{AFC14}
M.~Araujo, A.~Feix, F.~Costa, C.~Brukner, Quantum circuits cannot control unknown operations, New J. Phys. 16 (2014) 093026.

\bibitem{OGH16}
M.~Oszmaniec, A.~Grudka, M.~Horodecki, A.~W\'ojcik, Creating a superposition of unknown quantum states, Phys. Rev. Lett. 116 (2016) 110403.

\bibitem{TMV18}
J.~Thompson, K.~Modi, V.~Vedral, M.~Gu, Quantum plug n’ play: modular computation in the quantum regime, New J. Phys. 20 (2018) 013004.

\bibitem{WST08}
S.~Wehner, C.~Schaffner, B.~M. Terhal, Cryptography from noisy storage, Phys. Rev. Lett. 100 (2008) 220502.

\bibitem{W22_qvn}
D.-S. Wang, {A prototype of quantum von Neumann architecture}, Commun. Theor. Phys. 74 (2022) 095103.

\bibitem{CDP08a}
G.~Chiribella, G.~M. D'Ariano, P.~Perinotti, Transforming quantum operations: Quantum supermaps, Europhys. Lett. 83 (2008) 30004.

\bibitem{LWLW23}
Y.-T. Liu, K.~Wang, Y.-D. Liu, D.-S. Wang, {A Survey of Universal Quantum von Neumann Architecture}, Entropy 25~(8) (2023) 1187.

\bibitem{Kra83}
K.~Kraus, States, Effects, and Operations: Fundamental Notions of Quantum Theory, Vol. 190 of Lecture Notes in Physics, Springer-Verlag, Berlin, 1983.

\bibitem{Cho75}
M.-D. Choi, Completely positive linear maps on complex matrices, Linear Algebra Appl. 10 (1975) 285--290.

\bibitem{Jam72}
A.~Jamio{\l}kowski, Linear transformations which preserve trace and positive semidefiniteness of operators, Rep. Math. Phys. 3 (1972) 275.

\bibitem{W15_QS}
D.-S. Wang, Weak, strong, and uniform quantum simulations, Phys. Rev. A 91 (2015) 012334.

\bibitem{YSM23}
S.~Yoshida, A.~Soeda, M.~Murao, Reversing unknown qubit-unitary operation, deterministically and exactly, Phys. Rev. Lett. 131 (2023) 120602.

\bibitem{MZC25}
Y.~Mo, L.~Zhang, Y.~A. Chen, et~al., Parameterized quantum comb and simpler circuits for reversing unknown qubit-unitary operations, npj Quantum Inf. 11 (2025) 32.

\bibitem{Aha03}
D.~Aharonov, A simple proof that toffoli and hadamard are quantum universal, arXiv preprint arXiv:0301040 (2003).

\bibitem{BBC+93}
C.~H. Bennett, G.~Brassard, C.~Cr\'epeau, R.~Jozsa, A.~Peres, W.~K. Wootters, Teleporting an unknown quantum state via dual classical and einstein-podolsky-rosen channels, Phys. Rev. Lett. 70 (1993) 1895--1899.

\bibitem{GC99}
D.~Gottesman, I.~L. Chuang, Demonstrating the viability of universal quantum computation using teleportation and single-qubit operations, Nature 402~(6760) (1999) 390--393.

\bibitem{BK05}
S.~Bravyi, A.~Kitaev, {Universal quantum computation with ideal Clifford gates and noisy ancillas}, Phys. Rev. A 71 (2005) 022316.

\bibitem{KL98}
E.~Knill, R.~Laflamme, Power of one bit of quantum information, Phys. Rev. Lett. 81 (1998) 5672--5675.

\bibitem{KSV02}
A.~Kitaev, A.~H. Shen, M.~N. Vyalyi, Classical and Quantum Computation, Vol.~47 of Graduate Studies in Mathematics, American Mathematical Society, Providence, 2002.

\bibitem{Sho94}
P.~W. Shor, Algorithms for quantum computation: discrete logarithms and factoring, in: Proceedings 35th annual symposium on foundations of computer science, IEEE, 1994, pp. 124--134.

\bibitem{BCW+01}
H.~Buhrman, R.~Cleve, J.~Watrous, R.~de~Wolf, Quantum fingerprinting, Phys. Rev. Lett. 87 (2001) 167902.

\bibitem{BHM02}
G.~Brassard, P.~Hoyer, M.~Mosca, A.~Tapp, Quantum amplitude amplification and estimation, Contem. Mathemat. 305 (2002) 53–74.

\bibitem{GSLW19}
A.~Gilyen, Y.~Su, G.~H. Low, N.~Wiebe, Quantum singular value transformation and beyond: exponential improvements for quantum matrix arithmetics, in: Proceedings of the 51st Annual ACM SIGACT Symposium on Theory of Computing, 2019.

\bibitem{Long11}
G.~L. Long, Duality quantum computing and duality quantum information processing, Int. J. Theor. Phys. 50 (2011) 1305.

\bibitem{BCC+15}
D.~W. Berry, A.~M. Childs, R.~Cleve, R.~Kothari, R.~D. Somma, Simulating hamiltonian dynamics with a truncated taylor series, Phys. Rev. Lett. 114 (2015) 090502.

\bibitem{LKS+19}
H.~Levine, A.~Keesling, G.~Semeghini, A.~Omran, T.~T. Wang, S.~Ebadi, H.~Bernien, M.~Greiner, V.~Vuleti\ifmmode~\acute{c}\else \'{c}\fi{}, H.~Pichler, M.~D. Lukin, Parallel implementation of high-fidelity multiqubit gates with neutral atoms, Phys. Rev. Lett. 123 (2019) 170503.

\bibitem{KM20}
M.~Khazali, K.~M\o{}lmer, Fast multiqubit gates by adiabatic evolution in interacting excited-state manifolds of rydberg atoms and superconducting circuits, Phys. Rev. X 10 (2020) 021054.

\bibitem{KMN+22}
Y.~Kim, A.~Morvan, L.~B. Nguyen, R.~K. Naik, C.~Jünger, L.~Chen, J.~M. Kreikebaum, D.~I. Santiago, I.~Siddiqi, High-fidelity three-qubit itoffoli gate for fixed-frequency superconducting qubits, Nat. Phys. 18 (2022) 783.

\bibitem{BBC+95}
A.~Barenco, C.~H. Bennett, R.~Cleve, D.~P. DiVincenzo, N.~Margolus, P.~Shor, T.~Sleator, J.~A. Smolin, H.~Weinfurter, Elementary gates for quantum computation, Phys. Rev. A 52~(5) (1995) 3457.

\bibitem{SM13}
M.~Saeedi, I.~Markov, Synthesis and optimization of reversible circuits-a survey, ACM Comput. Surv. 45 (2013) 21.

\bibitem{ICK+16}
R.~Iten, R.~Colbeck, I.~Kukuljan, J.~Home, M.~Christandl, Quantum circuits for isometries, Phys. Rev. A 93 (2016) 032318.

\bibitem{BLP16}
B.~J. Brown, D.~Loss, J.~K. Pachos, C.~N. Self, J.~R. Wootton, Quantum memories at finite temperature, Rev. Mod. Phys. 88 (2016) 045005.

\bibitem{SB25}
S.~Stastny, G.~Burkard, The singlet-triplet and exchange-only flopping-mode spin qubits, arXiv preprint arXiv:2503.05032 (2025).

\bibitem{BEH23}
W.~C. Burton, B.~Estey, I.~M. Hoffman, A.~R. Perry, C.~Volin, G.~Price, Transport of multispecies ion crystals through a junction in a radio-frequency paul trap, Phys. Rev. Lett. 130 (2023) 173202.

\bibitem{BEG24}
D.~Bluvstein, S.~J. Evered, A.~A. Geim, et~al., Logical quantum processor based on reconfigurable atom arrays, Nature 626 (2024) 58.

\bibitem{RB01}
R.~Raussendorf, H.~J. Briegel, A one-way quantum computer, Phys. Rev. Lett. 86 (2001) 5188--5191.

\bibitem{KLM01}
E.~Knill, R.~Laflamme, G.~Milburn, A scheme for efficient quantum computation with linear optics, Nature 409 (2001) 46.

\bibitem{BR05}
D.~E. Browne, T.~Rudolph, Resource-efficient linear optical quantum computation, Phys. Rev. Lett. 95 (2005) 010501.

\bibitem{Abu24}
M.~AbuGhanem, Photonic quantum computers, arXiv preprint arXiv:2409.08229 (2024).

\bibitem{RZB94}
M.~Reck, A.~Zeilinger, H.~J. Bernstein, P.~Bertani, Experimental realization of any discrete unitary operator, Phys. Rev. Lett. 73 (1994) 58--61.

\bibitem{ZLC00}
X.~Zhou, D.~W. Leung, I.~L. Chuang, Methodology for quantum logic gate construction, Phys. Rev. A 62 (2000) 052316.

\bibitem{HOM87}
C.~K. Hong, Z.~Y. Ou, L.~Mandel, Measurement of subpicosecond time intervals between two photons by interference, Phys. Rev. Lett. 59 (1987) 2044--2046.

\bibitem{ZZH97}
M.~\ifmmode~\dot{Z}\else \.{Z}\fi{}ukowski, A.~Zeilinger, M.~A. Horne, Realizable higher-dimensional two-particle entanglements via multiport beam splitters, Phys. Rev. A 55 (1997) 2564--2579.

\bibitem{TTM10}
M.~C. Tichy, M.~Tiersch, F.~de~Melo, F.~Mintert, A.~Buchleitner, Zero-transmission law for multiport beam splitters, Phys. Rev. Lett. 104 (2010) 220405.

\bibitem{CAP08}
G.~Chiribella, G.~M. D'Ariano, P.~Perinotti, Memory effects in quantum channel discrimination, Phys. Rev. Lett. 101 (2008) 180501.

\bibitem{Hay22}
M.~Hayashi, Oblivious quantum computation and delegated multiparty quantum computation, arXiv preprint arXiv:2211.00962 (2022).

\bibitem{MSG+22}
J.~Morris, V.~Saggio, A.~Gocanin, B.~Dakic, Quantum verification and estimation with few copies, Adv. Quantum Technol. 5 (2022) 2100118.

\bibitem{CGS02}
C.~Cr{\'e}peau, D.~Gottesman, A.~Smith, Secure multi-party quantum computation, in: In STOC ’02: Proc. 34rd Annual ACM Symp. Theory of Computing, 2002, p. 643.

\bibitem{SLK20}
A.~Sebastian, M.~Le~Gallo, R.~Khaddam-Aljameh, et~al., Memory devices and applications for in-memory computing, Nat. Nanotechnol. 15 (2020) 529.

\bibitem{WW23}
K.~Wang, D.-S. Wang, Quantum circuit simulation of superchannels, New J. Phys. 25~(4) (2023) 043013.

\bibitem{MPS18}
K.~Modi, A.~K. Pati, A.~Sen(De), U.~Sen, Masking quantum information is impossible, Phys. Rev. Lett. 120 (2018) 230501.

\bibitem{Wat18}
J.~Watrous, {The Theory of Quantum Information}, Cambridge University Press, 2018.

\end{thebibliography}
\bibliographystyle{elsarticle-num}

\end{document}